\documentclass[12pt,thmsa,arabtex]{article}
\usepackage{amsfonts}
\usepackage{graphicx}

\input{tcilatex}
\begin{document}

\begin{center}
\bigskip

{\Large Generation of three-qubit entangled states using coupled
multi-quantum dots }

M. Abdel-Aty$^{1}$, M. R. B. Wahiddin$^{1}$ and A.-S. F. Obada$^{2}$

$^{1}${\small Cyberspace Security Laboratory, MIMOS Berhad, Technology Park,
57000 Kuala Lumpur, Malaysia}

$^{2}${\small Mathematics Department, Faculty of Science, Al-Azhar
University, Naser City, Cairo, Egypt}
\end{center}

We discuss a mechanism for generating a maximum entangled state (GHZ) in a
coupled quantum dots system, based on analytical techniques. The reliable
generation of such states is crucial for implementing solid-state based
quantum information schemes. The signature originates from a remarkably weak
field pulse or a far off-resonance effects which could be implemented using
technology that is currently being developed. The results are illustrated
with an application to a specific wide-gap semiconductor quantum dots
system, like Zinc Selenide (ZnSe) based quantum dots.


\section{Introduction}

Excitons (electron-hole bound states) within quantum dots (QDs) have
attracted much interest in the field of quantum computation and have formed
the basis of several proposals for quantum logic gates \textrm{\cite{naz05}}%
. The energy shift due to the exciton-exciton dipole interaction between
multi-quantum dots gives rise to diagonal terms in the interaction
Hamiltonian, and hence it has been proposed that quantum logic may be
performed via ultra-fast laser pulses \cite{bra05}. However, excitons within
adjacent quantum dots are also able to interact through their resonant (F%
\"{o}rster) energy transfer, some evidence for which has been obtained
experimentally \cite{ber02}. Because the quantum dot has discrete energy
levels, much like an atom, the energy levels can be controlled by changing
the size and shape of the quantum dot, and the depth of the potential. Like
in atoms, the energy levels of small quantum dots can be probed by optical
spectroscopy techniques. In contrast to atoms it is relatively easy to
connect quantum dots by tunnel barriers to conducting leads, which allows
the application of the techniques of tunneling spectroscopy for their
investigation.

Many properties of such systems can be investigated by transport, if the
dots are fabricated between contacts acting as source and drain for
electrons which can enter or leave the dot \cite{pas06,pas06b}. Creating
entangled states is the first step toward studying any effects related to
entanglement. Semiconductor quantum dots have their own advantages as a
candidate of the basic building blocks of solid-state-based quantum logic
devices \cite{los98,ste06,yi01}, due to the existence of an industrial base
for semiconductor processing and due to the ease of integration with
existing devices. The experimental realization of optically induced
entanglement of excitons in a single quantum dot \cite{che00} and
theoretical study on coupled quantum dots \cite{rei00} were reported most
recently. In those investigations a classical laser field is applied to
create the electron-hole pair in the dot(s).

The issue we have in mind has to do with the exact solution of the system,
taking into account the dependence on the relevant magnitudes such as the
field strength, inter-dot process hoping rate and the detuning. This is most
conveniently accomplished in a quantum formalism in terms of the Schr\"{o}%
dinger equation. Related treatments based on either adiabatic elimination 
\cite{pas06}, discussing entangled state generation conditions, or the
coupled equations without the detuning dependence, have been presented in
the literature \cite{los98}. What we have studied and present below is
essentially the most general case of the complete system equations. To be
more precise, with this approach we could create maximum entangled states
without using the approximation methods adapted in previous studies.

The outline of this paper is arranged as follows: in section 2, we give
notations and definitions of the model followed by a rigorous analytical
approach for obtaining exact-time dependent expressions for the probability
amplitudes. Section 3\textbf{\ }is devoted to consider the maximum entangled
state generation. By a numerical computation, we examine the influence of
different parameters on the evolution of the probability density of finding
the maximum entangled state. Finally, our conclusion will be presented in
section 4\textbf{.}

\section{Model}

We characterize the electron and hole states within a quantum dots model, as
well as accounting for the binding energy due to electron-hole coupling
within a dot when estimating the ground state exciton energy. To begin with,
we consider a system with $3$ identical quantum dots that are coupled by the
F\"{o}rster process \cite{ola06}. This process originates from the Coulomb
interaction whereby an exciton can hop between the dots \cite{nie00}. 
\textrm{The Coulomb exchange interaction in QDs gives rise to a
non-radiative resonant energy transfer (i.e. F\"{o}rster process) which
corresponds to the exchange of a virtual photon, thereby destroying an
exciton in a dot and then re-creating it in a close by dot (for a detailed
discussion of how to exploit the F\"{o}rster interaction see e.g. \cite%
{rei00,qui99}). }The present study is motivated by recent experimental
results which demonstrated the optical detection of an NMR signal in single
QDs \cite{bro98}. Hence the underlying nuclear spins in the QDs can indeed
be controlled with optical techniques, via the electron-nucleus coupling.
Also, few electron dots can be prepared experimentally \cite{ash93} and
their magic number transitions could be measured as a function of magnetic
field.

Keeping in mind the fact that all constant energy terms may be ignored, the
total Hamiltonian governs the time evolution of single excitons within the
individual quantum-dot systems with its interdot F\"{o}rster hopping and the
interaction of the system with the laser field, in an appropriate rotating
frame \cite{rod01}, is given by 
\begin{equation}
\hat{H}=\hat{H}_{0}+\hat{H}_{F}+\hat{H}_{I},  \label{ham}
\end{equation}%
where, $\hat{H}_{0}$ is the single-exciton Hamiltonian, $\hat{H}_{F}$ is the
interdot F\"{o}rster interaction and $\hat{H}_{I}$ is the coupling of the
carrier system with a classical laser field, which can be written as 
\begin{eqnarray}
\hat{H}_{0} &=&\frac{\varepsilon }{2}\sum\limits_{j=1}^{3}\left( \widehat{e}%
_{j}^{\dagger }\widehat{e}_{j}-\widehat{\psi }_{j}\widehat{\psi }%
_{j}^{\dagger }\right) ,  \nonumber \\
\hat{H}_{F} &=&\frac{\hbar \eta }{2}\sum\limits_{j=1}^{3}\sum%
\limits_{k=1}^{3}\left( \widehat{e}_{j}^{\dagger }\widehat{\psi }_{k}%
\widehat{e}_{k}\widehat{\psi }_{j}^{\dagger }+\widehat{\psi }_{j}\widehat{e}%
_{k}^{\dagger }\widehat{\psi }_{k}^{\dagger }\widehat{e}_{j}\right) , 
\nonumber \\
\hat{H}_{I} &=&\frac{\hbar }{2}\sum\limits_{j=1}^{3}\left( \Omega \exp \left[
-i\omega t+i\phi \right] \widehat{e}_{j}^{\dagger }\widehat{\psi }%
_{j}^{\dagger }+\Omega ^{\ast }\exp \left[ i\omega t-i\phi \right] \widehat{%
\psi }_{j}\widehat{e}_{j}\right) 
\end{eqnarray}%
where $\Omega $ represents the Rabi frequency which characterizes the
laser-quantum dot coupling and $\hbar \Omega =\mu E$, where $\mu $ is the
coupling strength. We denote by $E$, $\phi $ the laser pulse electric field
amplitude and phase, respectively. The parameter $\omega $ discribes the
angular frequency of the laser field. The operator $\widehat{e}_{j}(\widehat{%
\psi }_{j})$ is the electron (hole) annihilation operator and $\widehat{e}%
_{j}^{\dagger }(\widehat{\psi }_{j}^{\dagger })$ is the electron (hole)
creation operator in the $j^{\underline{th}}$ quantum dot. The parameter $%
\varepsilon $ describes the band gap energy of the quantum dot and $\eta $
is the inter-dot process hoping rate.

\bigskip

For a coupled three-quantum-dot system, we consider the $J=3/2$-subspace as
the only one optically active (the other $J=1/2$ subspace remains optically
dark). We work in the basis set $|J=3/2,M\rangle ,\{|0\rangle
=|3/2,-3/2\rangle ,|1\rangle =|3/2,-1/2\rangle ,|2\rangle =|3/2,1/2\rangle
,|3\rangle =|3/2,3/2\rangle \}$, where $|0\rangle $ is the vacuum state, $%
|1\rangle $ is the single-exciton state, $|2\rangle $ is the biexciton state
and $|3\rangle $ is the triexciton state. More specifically, applying the
rotating wave approximation and a unitary transformation, the resulting
Hamiltonian may be written as 
\begin{eqnarray}
H &=&\frac{3}{2}\left( \eta -\Delta \right) \widehat{S}_{11}+\Omega ^{\ast }%
\sqrt{3}e^{-i\phi }\widehat{S}_{12}+\Omega \sqrt{3}e^{i\phi }\widehat{S}%
_{21}+\frac{1}{2}\left( 7\eta -\Delta \right) \widehat{S}_{22}+2\Omega
^{\ast }e^{-i\phi }\widehat{S}_{23}  \nonumber \\
&&+2\Omega e^{i\phi }\widehat{S}_{32}+\frac{1}{2}\left( 7\eta +\Delta
\right) \widehat{S}_{33}+\Omega ^{\ast }\sqrt{3}e^{-i\phi }\widehat{S}%
_{34}+\Omega \sqrt{3}e^{i\phi }\widehat{S}_{43}+\frac{3}{2}\left( \eta
+\Delta \right) \widehat{S}_{44},  \label{h2}
\end{eqnarray}%
where $\widehat{S}_{ij}=|i+1\rangle \langle |j+1|$ are related to the above
states $|0\rangle ,$ $|1\rangle ,$ $|2\rangle $ and $|3\rangle .$ We denote
by $\Delta $ the detuning of the laser pulse from exact resonance ($\hbar
\Delta =\varepsilon -\hbar \omega )$. We consider the situation of a laser
pulse with central frequency $\omega $ given by $\xi (t)=\Omega e^{-i\omega
t}.$ From a practical point of view, parameters $\Omega $ and $\omega $ are
adjustable in the experiment to give control over the system of QDs.

We devote the following discussion to find an explicit expression for the
wave function in Schr\"{o}dinger picture. We use an analytic approach that
seeks to reduce the coupled differential equations (probability amplitudes)
to a solvable linear equation in order to study in detail the related
phenomena. To reach our goal we assume that the wave function of the
complete system may be expanded in terms of the eigenstates, namely 
\begin{equation}
\left\vert \Psi (t)\right\rangle =B_{0}(t)\left\vert 0\right\rangle
+B_{1}(t)\left\vert 1\right\rangle +B_{2}(t)\left\vert 2\right\rangle
+B_{3}(t)\left\vert 3\right\rangle .  \label{weq}
\end{equation}%
The time dependence of the amplitudes in equation (\ref{weq}) is governed by
the Schr\"{o}dinger equation with the Hamiltonian given by equation (\ref{h2}%
). In order to find the probability amplitudes $B_{i}(t)$, one may introduce
the function \cite{gui03} 
\begin{equation}
G(t)=B_{0}(t)+xB_{1}(t)+yB_{2}(t)+zB_{3}(t),
\end{equation}%
which leads to the following equation 
\begin{equation}
i\frac{dG(t)}{dt}=\beta \left( B_{0}(t)+\frac{\gamma _{1}}{\beta }B_{1}(t)+%
\frac{\gamma _{2}}{\beta }B_{2}(t)+\frac{\gamma _{3}}{\beta }B_{3}(t)\right)
.
\end{equation}%
where $\beta =\xi _{00}+x\xi _{10},$ $\gamma _{1}=\xi _{01}+x\xi _{11}+y\xi
_{21},{\ \ }\gamma _{2}=x\xi _{12}+y\xi _{22}+z\xi _{32},$ $\gamma _{3}=y\xi
_{23}+z\xi _{33},$ and $\xi _{jk}=\left\langle j\right\vert \hat{H}%
\left\vert k\right\rangle .$ In this case and using equations (\ref{h2}) and
(\ref{weq}), we have $\xi _{00}=1.5(\eta -\Delta ),\xi _{01}=\left( \xi
_{10}\right) ^{\ast }$ $=\sqrt{3}\Omega \exp (i\phi ),$ $\xi _{11}=0.5(7\eta
-\Delta ),$ $\xi _{12}=\left( \xi _{21}\right) ^{\ast }$ $=2\Omega \exp
(-i\phi ),$ $\xi _{22}=0.5(7\eta +\Delta ),$ $\xi _{23}=\left( \xi
_{32}\right) ^{\ast }$ $=\sqrt{3}\Omega \exp (-i\phi ),$ $\xi _{33}=1.5(\eta
+\Delta )$\ and $\xi _{ij}=0$ otherwise. Now let us seek a solution of $G(t)$
such that $\dot{G}(t)=-i\mu G(t)$. This holds if and only if $\mu =\beta
,x=\gamma _{1}/\beta ,$ $\ y=\gamma _{2}/\beta $ and $\ z=\gamma _{3}/\beta
. $ Therefore, after some minor algebraic calculations, the general solution
can be written as 
\begin{equation}
B_{j-1}(t)=\exp \left( -i\xi _{11}t\right) \sum\limits_{k=1}^{4}\lambda
_{jk}\exp \left( -i\xi _{21}\mu _{k}t\right) ,{\ \ \ \ }j=1,2,3,4
\label{amp}
\end{equation}%
where 
\begin{equation}
\mu _{1,2}=-\frac{\Re _{1}}{4}-\Upsilon _{1}\pm \left( \Upsilon _{2}-\frac{%
m_{4}}{m_{2}}\right) ^{\frac{1}{2}},{\ \ \ \ \ }\mu _{3,4}=-\frac{\Re _{1}}{4%
}+\Upsilon _{1}\pm \left( \Upsilon _{2}+\frac{m_{4}}{m_{2}}\right) ^{\frac{1%
}{2}},
\end{equation}%
and $\Upsilon _{1}=\frac{1}{2}\sqrt{\Re _{1}^{2}/4-2\Re
_{2}/3+m_{2}/(3m_{1})+m_{1}/(3\sqrt[3]{2})},$ $\Upsilon _{2}=\Re
_{1}^{2}/2-4\Re _{2}/3-m_{3}/(3m_{1})-m_{1}/(3\sqrt[3]{2}),$ $m_{1}=\left(
m_{0}+\sqrt{-2m_{3}^{2}+m_{0}^{2}}\right) ^{1/3},$ $m_{0}=2\Re _{2}^{2}-9\Re
_{1}\Re _{2}\Re _{3}+27\Re _{3}^{2}+27\Re _{1}^{2}\Re _{4}-72\Re _{2}\Re
_{4},$ $m_{2}=\sqrt[3]{2}\left( \Re _{2}^{2}-3\Re _{1}\Re _{3}+12\Re
_{4}\right) ,$ $m_{3}=4(\Re _{1}^{2}/4-2\Re _{2}/3$ $+m_{3}/(3m_{1})+m_{1}/(3%
\sqrt[3]{2})),$ $m_{4}=-\Re _{1}^{3}+4\Re _{1}\Re _{2}-8\Re _{3}.$ The
parameters $\Re _{i},(i=1,2,3,4)$ are the coefficients of the fourth order
equation for $x$ and $\lambda _{ij}=O_{ij}^{-1}G_{j}(0)$.

As a specific dynamical example, we discuss a frequently encountered
phenomena of particular interest in which we discuss the generation of
maximum entangled GHZ states in the present system.

\section{Greenberger--Horne--Zeilinger (GHZ)}

There is a class of genuine tripartite entanglement, that is the
Greenberger--Horne--Zeilinger (GHZ) state \cite{ger89,sai06} which is given
as

\begin{equation}
|GHZ\rangle _{\tau }=\left( \frac{1}{\sqrt{2}}|000\rangle +\frac{\exp (i\tau
)}{\sqrt{2}}|111\rangle \right) ,  \label{ghz}
\end{equation}%
for arbitrary values of $\tau $. Starting with a zero-exciton state as an
initial state, the probability density $\wp (GHZ)$ of finding the entangled $%
|GHZ\rangle _{\tau }$ state between vacuum and triexciton states can be
calculated in the following form

\begin{eqnarray}
\wp (GHZ) &=&\left\vert _{\tau }\left\langle GHZ|\Psi (t)\right\rangle
\right\vert ^{2}  \nonumber \\
&=&\frac{1}{2}\left\vert B_{0}(t)+B_{3}(t)\right\vert ^{2},
\end{eqnarray}%
where $B_{0}(t)$ and $B_{3}(t)$ are given by equation (\ref{amp}).

\begin{figure}[tbph]
\begin{center}
\includegraphics[width=13cm]{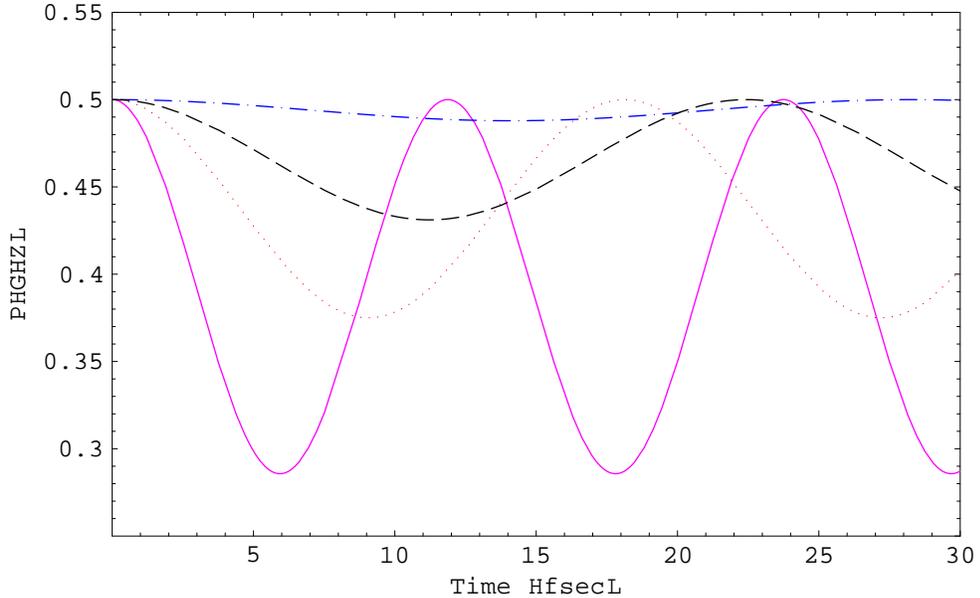}
\end{center}
\caption{Generation of the GHZ state $\frac{1}{\protect\sqrt{2}}|000\rangle +%
\frac{\exp (i\protect\tau )}{\protect\sqrt{2}}|111\rangle $. The parameters
are $\protect\eta =0.1,$ $\Delta =0,\protect\phi =0$ and for different
values of the laser-quantum dot coupling $\Omega =0.1$ $fs^{-1}$ (solid
curve), $\Omega =0.05$ $fs^{-1}$ (dotted curve), $\Omega =0.03$ $fs^{-1}$
(dashed curve), and $\Omega =0.01$ $fs^{-1}$ (dotted-dashed curve). }
\end{figure}
Here we present the results of numerical calculations for a specific
wide-gap semiconductor quantum dots system, like Zinc Selenide (ZnSe) based
quantum dots, where femtosecond spectroscopy is currently available for
these systems \cite{bar98}. For these materials, the band gap $\epsilon =2.8$
$eV$, which implies a resonant optical frequency $\omega =4.3\times
10^{15}s^{-1}$.

Specifically, we consider the probability density $\wp (GHZ)$ as a function
of the scaled time and different values of the laser-quantum dot coupling in
figure 1 and different values of the detuning parameter in figure 2. When we
consider the parameter $\Omega =0.1$ $fs^{-1},$ we see that the probability
density shows periodic oscillations between $2.6$ and $0.51$ (see figure 1).
Once the laser-quantum dot coupling is decreased, the amplitude of the
oscillations becomes smaller and smaller. Further decreasing of this
parameter, we obtain a maximally entangled GHZ state in the system of the
three coupled quantum dots i.e $\wp (GHZ)\approx 0.5.$ Also, from figure 1,
we see that the selective pulses used to create such maximally entangled GHZ
states in the present system is $\pi /2$. The generation of the GHZ state $%
|GHZ\rangle _{\tau }$ requires a time $1.3\times 10^{-14}$ sec and explore
several different ranges for the time pulses required in the generation of
such GHZ states.

\begin{figure}[tbph]
\begin{center}
\includegraphics[width=13cm]{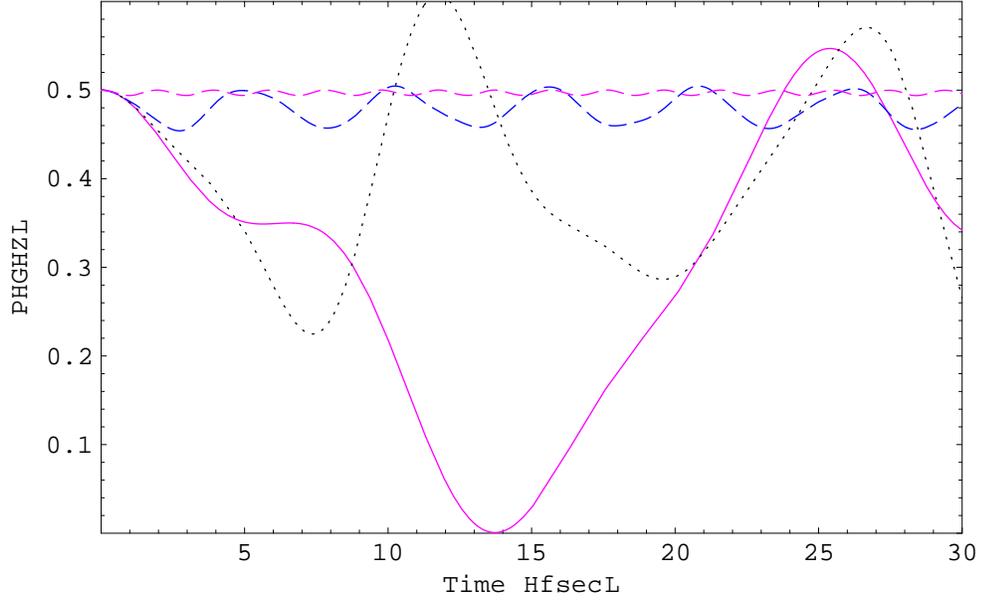}
\end{center}
\caption{Generation of the GHZ state $\frac{1}{\protect\sqrt{2}}|000\rangle +%
\frac{\exp (i\protect\tau )}{\protect\sqrt{2}}|111\rangle $. The parameters
are $\protect\eta =0.1,$ $\Omega =0.05$ $fs^{-1},\protect\phi =0$ and for
different values of the detuning parameter, where $\Delta =0.1\protect\eta $
(solid curve), $\Delta =0.3\protect\eta $ (dotted curve), $\Delta =1.0%
\protect\eta $ (dashed curve), and $\Delta =3.0\protect\eta $ (dotted-dashed
curve). }
\end{figure}

However, interestingly, this does not mean that there is no other effect due
to other parameters of the system in this case. As we show next, there is an
effect due to far off-resonance interaction, even though reversing the sign
or changing the values of the parameter $\tau $ has no effect. The effect of
the detuning parameter on the probability density may be examined by
considering the far off-resonance interaction, namely, $\Delta >0.$
Although, we have found that it is not difficult to find maximum entangled
state using weak field limit (see figure 1), part of the motivation for this
paper grew out of an effort to obtain maximum entangled state using our
analytical solution, based on effects of the detuning parameter. This is
shown in figure 2, where we plot the probability density $\wp (GHZ)$ as a
function of the scaled time for different values of the detuning parameter.
All the parameters are identical to those used in figure 1, except for the
detuning parameter. As we show, the maximum entangled GHZ state is obtained
when the detuning parameter is increased enough \ even in the presence of
strong field limit (see figure 2). Therefore, the same maximum entangled GHZ
state is formed periodically. By choosing a suitable time interval and large
detuning, one can effectively obtain $\wp (GHZ)\approx 0.5.$ Now the
difference between the effects of the two cases is obvious. Then, the most
reliable way to obtain maximum entangled GHZ state is the consideration of
the weak field limit or far off-resonant interaction. In the above figures,
we have shown how maximally entangled GHZ states can be generated using the
optically driven resonant transfer of excitons between quantum dots. \ 
\textrm{Based on such sensitivity and some other evidence, we suspect that
the analytical results presented here, could be attained for a larger number
of particles. }

\section{Conclusion}

In summary, an analytical solution for single excitons within three
individual ZeSe quantum dots and their interdot hopping in the presence of
the F\"{o}rster interaction has been developed and discussed. Such
analytical solutions provide useful physical insight, which together with
numerical treatments are used to generate maximum entangled GHZ states. More
explicitly, in the exciton system, the large values of the detuning help in
generating maximum entangled states. \ Nevertheless, the calculations
indicate that the maximum entangled states can still exist, even for the
resonant case, when the electron and hole are driven by a suitable laser
field (weak field limit). This study reveals that the three coupled quantum
dots can be used for generating a maximum entangled states, such as GHZ.
Paramount importance is acquiring the maximum entangled states due to an
ensemble of three quantum dots as it is becoming ubiquitous to different
fields of application as quantum computation, quantum information processing
and other related fields.

{\large Acknowledgments:}

We are indebted to E. Paspalakis, M. Salman and S. S. Hassan for valuable
discussions which led to the investigation above.

\end{document}